\documentclass[preprintnumbers]{revtex4}

\usepackage{graphicx,epsfig}
\def\MPL #1 #2 #3 {Mod. Phys. Lett. {\bf#1},\ #2 (#3)}
\def\NPB #1 #2 #3 {Nucl. Phys. {\bf#1},\ #2 (#3)}
\def\PLB #1 #2 #3 {Phys. Lett. {\bf#1},\ #2 (#3)}
\def\PR #1 #2 #3 {Phys. Rep. {\bf#1},\ #2 (#3)}
\def\PRD #1 #2 #3 {Phys. Rev. {\bf#1},\ #2 (#3)}
\def\PRL #1 #2 #3 {Phys. Rev. Lett. {\bf#1},\ #2 (#3)}
\def\RMP #1 #2 #3 {Rev. Mod. Phys. {\bf#1},\ #2 (#3)}
\def\NIM #1 #2 #3 {Nuc. Inst. Meth. {\bf#1},\ #2 (#3)}
\def\ZPC #1 #2 #3 {Z. Phys. {\bf#1},\ #2 (#3)}
\def\EJPC #1 #2 #3 {E. Phys. J. {\bf#1},\ #2 (#3)}
\def\IJMP #1 #2 #3 {Int. J. Mod. Phys. {\bf#1},\ #2 (#3)}
\def\JHEP #1 #2 #3 {J. High En. Phys. {\bf#1},\ #2 (#3)}

\begin{document}
\bibliographystyle{revtex}

\preprint{SLAC-PUB-9082}

\title{\boldmath Graviton Production at CLIC}




\author{M. Battaglia}
\email[]{marco.battaglia@cern.ch}
\affiliation{CERN, Geneva, Switzerland}

\author{A. De Roeck}
\email[]{deroeck@mail.cern.ch}
\affiliation{CERN, Geneva, Switzerland}

\author{T. Rizzo}
\email[]{rizzo@slac.stanford.edu}
\affiliation{Stanford Linear Accelerator Center, 
Stanford University, Stanford, California 94309 USA}


\date{\today}


\begin{abstract}
\vspace*{0.25cm}
Direct production of Kaluza-Klein states in the TeV range is studied for
the experimental environment at the multi-TeV $e^+e^-$ collider CLIC.
The sensitivity of such data to  model parameters is discussed 
for  the Randall-Sundrum(RS) and TeV scale extra dimensional models.
\end{abstract}


\maketitle

\section{Introduction}

For the past few years the phenomenology of large (extra) dimensions has been 
explored at the TeV scale. These theories aim to solve the hierarchy problem
by bringing the Gravity scale  closer to the  Electroweak scale.
Extra dimension signatures at future colliders, particularly at
CLIC,  have been discussed extensively 
in~\cite{rizzo}. Here we report on  a  study using two models
which can produce measurable resonances in the 
two-fermion production cross section, 
in the multi-TeV range which would be accessible at CLIC.
CLIC is conceived to be an $e^+e^-$ linear collider (LC) optimized for 
a centre of mass (CMS) energy  of  3 TeV with 
 ${\cal L} \cong 10^{35}$cm$^{-2}$s$^{-1}$, using
a novel acceleration concept called two-beam acceleration~\cite{clic}.

For these studies
the  CLIC physics working group tools have been used, which include
 smearing of the CMS energy of the 
collisions according to the  CLIC luminosity spectrum~\cite{calypso}; 
 overlaying the  hadronic background from
$\gamma\gamma$ events~\cite{calypso}; and producing the
 detector response by  SIMDET, a fast simulation  package~\cite{simdet}.
 Events of the type 
$e^+e^- \rightarrow f\overline{f}$ have been generated with the 
PYTHIA event generator{\cite {pythia}}.

\section{Randall-Sundrum model}

In the extra-dimension scenario proposed by 
Randall-Sundrum(RS)~\cite{RS}  the hierarchy between the Planck 
and the Electroweak scale is generated by an exponential function called
'warp factor'. This model predicts Kaluza-Klein graviton resonances with both
weak scale masses and couplings to matter. Hence the production of 
TeV scale graviton resonances is  
expected  in two fermion channels{\cite {dhr}}. 
In its simplest version, with two branes and one 
extra dimension, and where all of the SM fields remain on the brane, 
the model has 
two fundamental parameters: the mass of the first Kaluza-Klein state, $m_1$, 
and the parameter $c= k/M_{\overline{Pl}}$, where $k$ is related to the 
curvature of the 5-dimensional space and $M_{\overline{Pl}}$ is the 4-d 
effective Planck scale. The parameter $c$
controls the effective coupling
strength of the graviton and thus the width of the resonances, and 
should be less than one but yet not too far away from unity. 
\begin{figure}[htbp]   
\begin{center} 
\epsfig{file=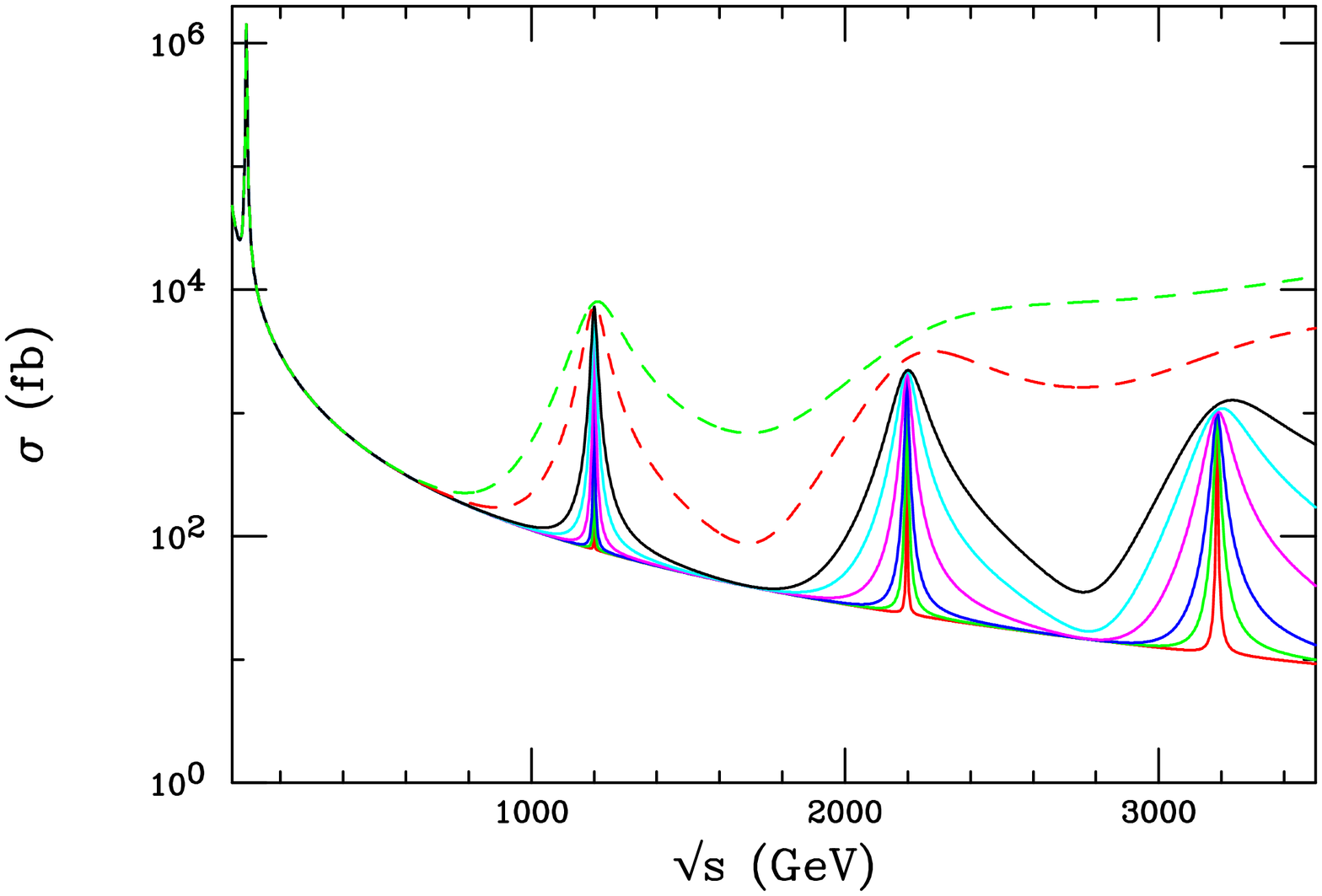,bbllx=90,bblly=90,bburx=550,bbury=720,width=7cm,angle=90,clip}
\epsfig{file=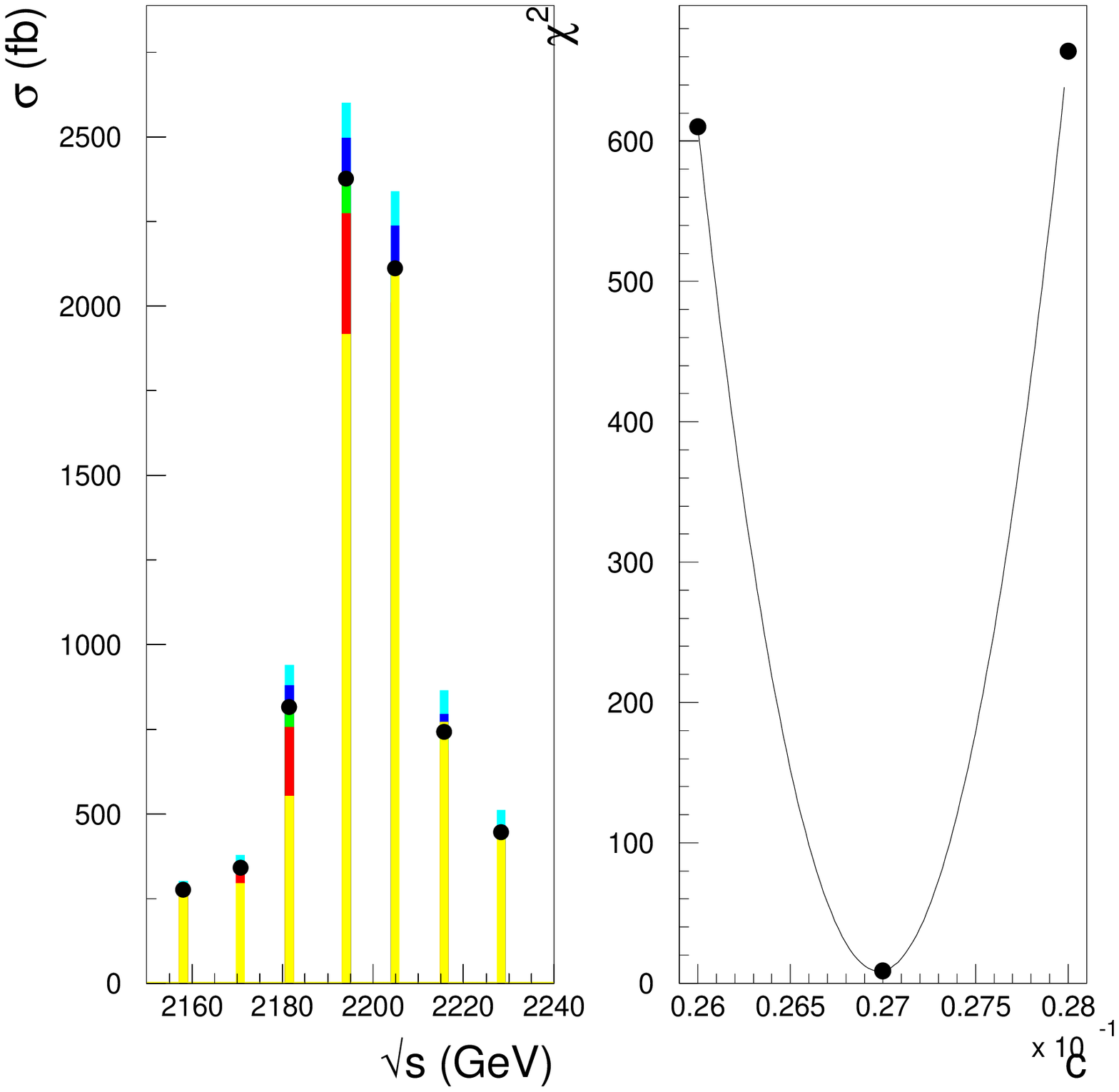,bbllx=20,bblly=140,bburx=540,bbury=690,width=6.5cm}
\caption{
(Left) KK graviton excitations in the RS model 
produced in the process $e^+e^-\to \mu^+\mu^-$. From the most narrow to widest 
resonances the curves are for $c$ in the range 0.01 to 0.2. 
(Right) Scan of the resonance (7 points) and $\chi^2$ fit to the
measured spectrum for models with different values for $c$.
}

\label{fig:kk1}
\end{center}
\end{figure}

The resulting  spectrum for $e^+e^-\rightarrow \mu^+\mu^-$ is shown in 
Fig.~\ref{fig:kk1}. The cross sections are huge and the 
signal cannot be missed at a LC with sufficient CMS  energy.
If such resonances are observed -- perhaps first by the LHC in the 
range of a few TeV-- it will be important to establish the nature of these
newly produced particles, i.e. to measure  
their properties (mass, width and branching ratios) and quantum numbers (spin).
Note that  the mass $m_1$ of the first resonance determines the 
resonance pattern: the masses of all higher mass resonances are then fixed.

The signal for one KK resonance ($G_1$) is implemented in PYTHIA 
6.158~\cite{pythia}
via process 41.  For this study PYHTIA has been extended 
to include  two more resonances
($G_2, G_3$,  corresponding to processes 42 and 43) 
to allow to check the measurability of the graviton self-coupling.
The decay branching ratios  of these
resonances were modified according to~\cite{dhr,rizzo3}. 
In particular the gravitons can decay into two photons in about
4\% of the cases, a signature which would distinguish them from e.g. 
new heavy $Z'$ states{\cite {snow}}.

\begin{figure}[htbp]   
\begin{center} 
\epsfig{file=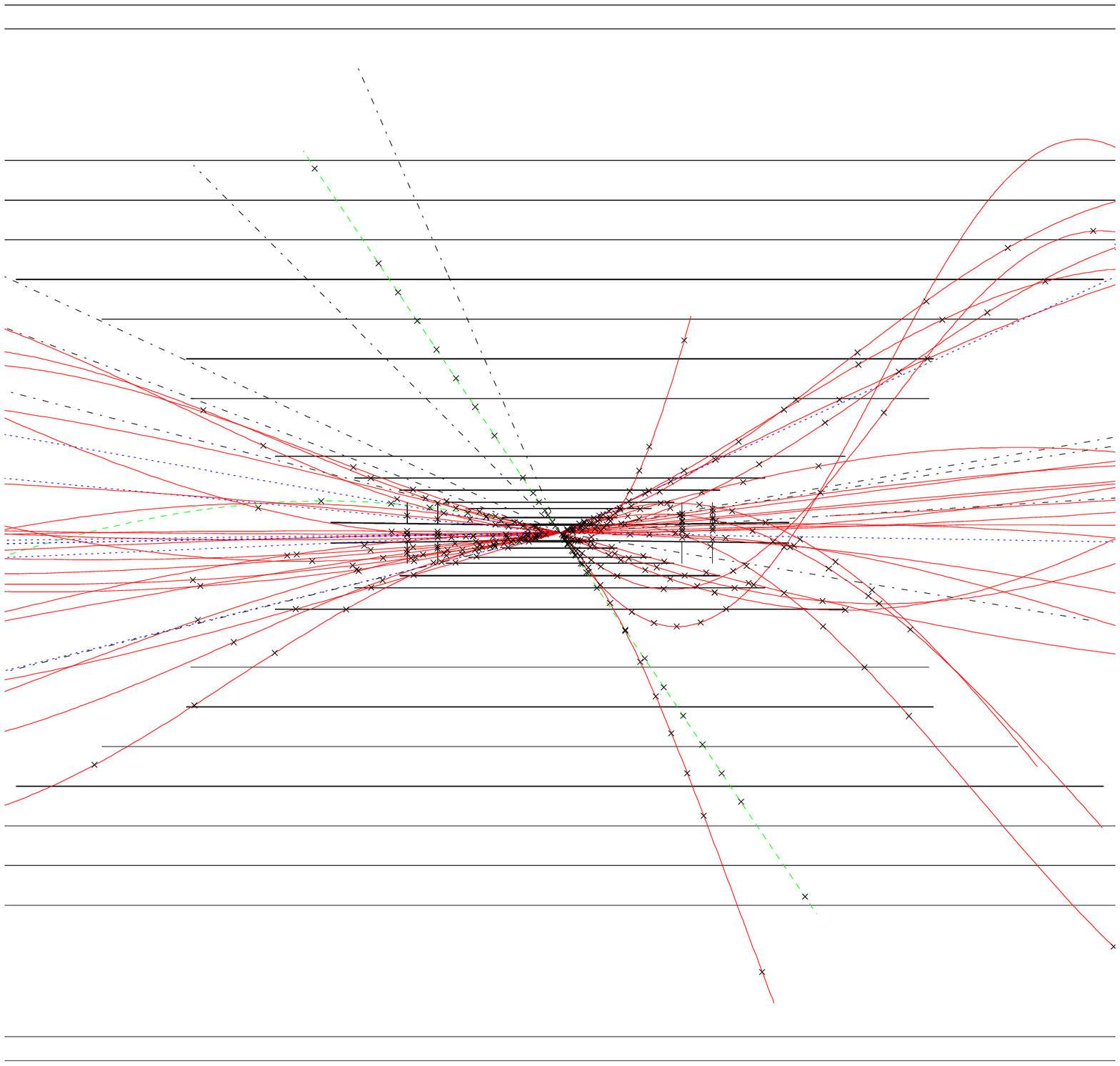,bbllx=0,bblly=100,bburx=600,bbury=700,width=6cm}
\epsfig{file=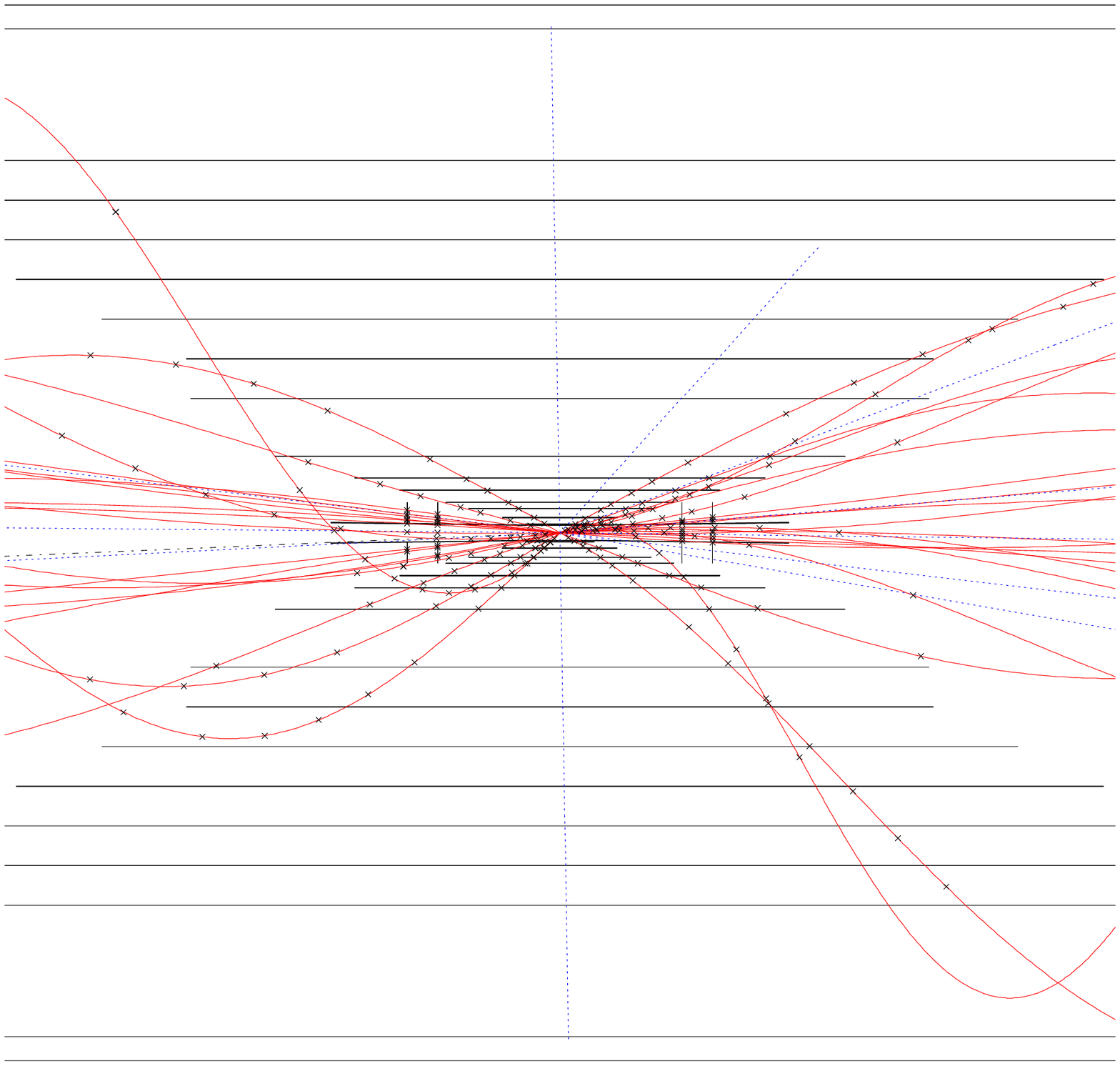,bbllx=0,bblly=100,bburx=600,bbury=700,width=6cm}
\caption{(Left) $G_3 (3200\, \rm{GeV})\rightarrow \mu\mu$ decay and 
(Right) $G_3 (3200\,  \rm{GeV})\rightarrow \gamma\gamma$ decay in a detector
at CLIC, with $\gamma\gamma$ background events overlaid
(one bunch crossing). }
\label{fig:kk2}
\end{center}
\end{figure}

\begin{figure}[htbp]   
\begin{center} 
\epsfig{file=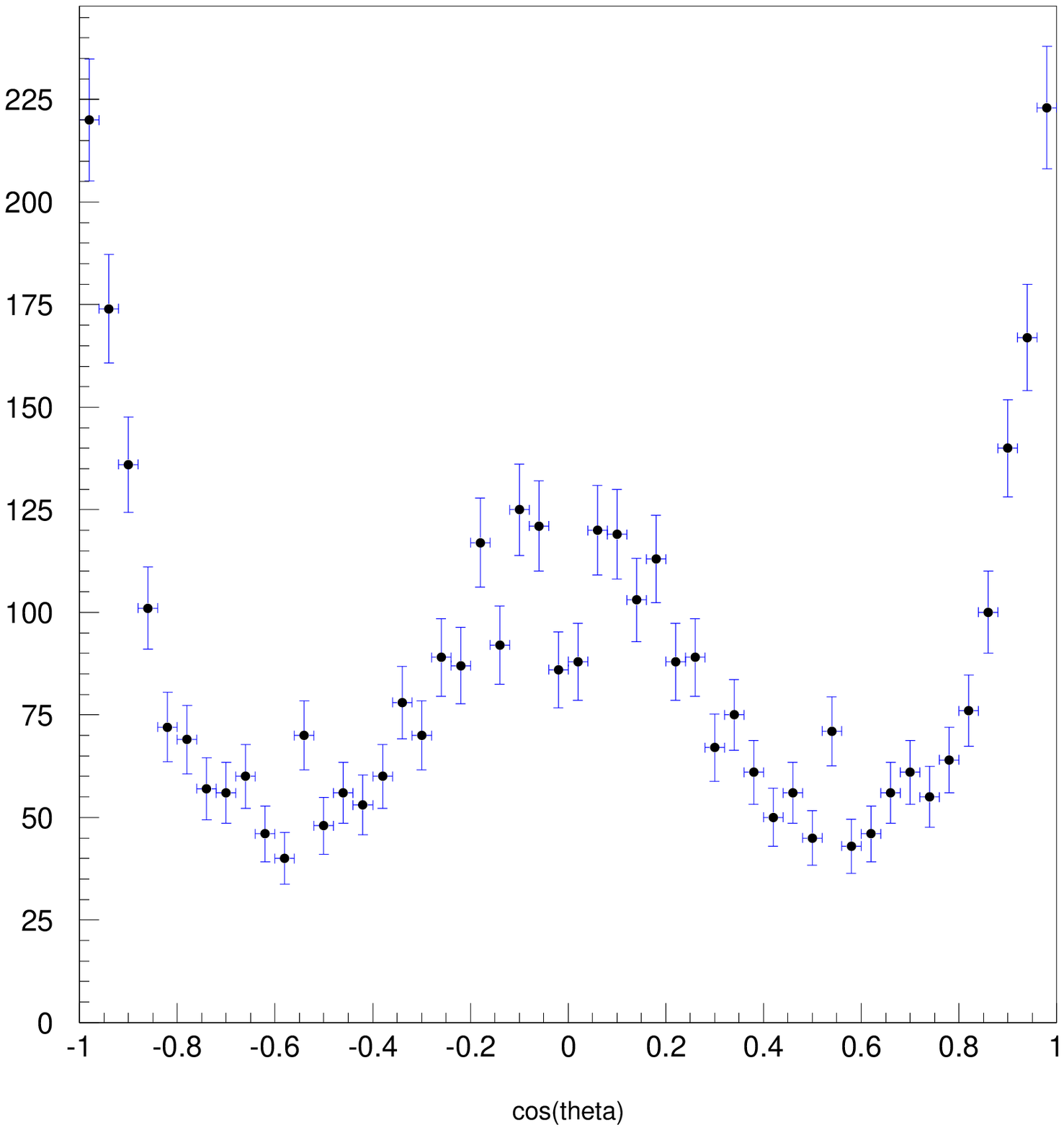,bbllx=0,bblly=0,bburx=565,bbury=565,width=7.5cm}
\epsfig{file=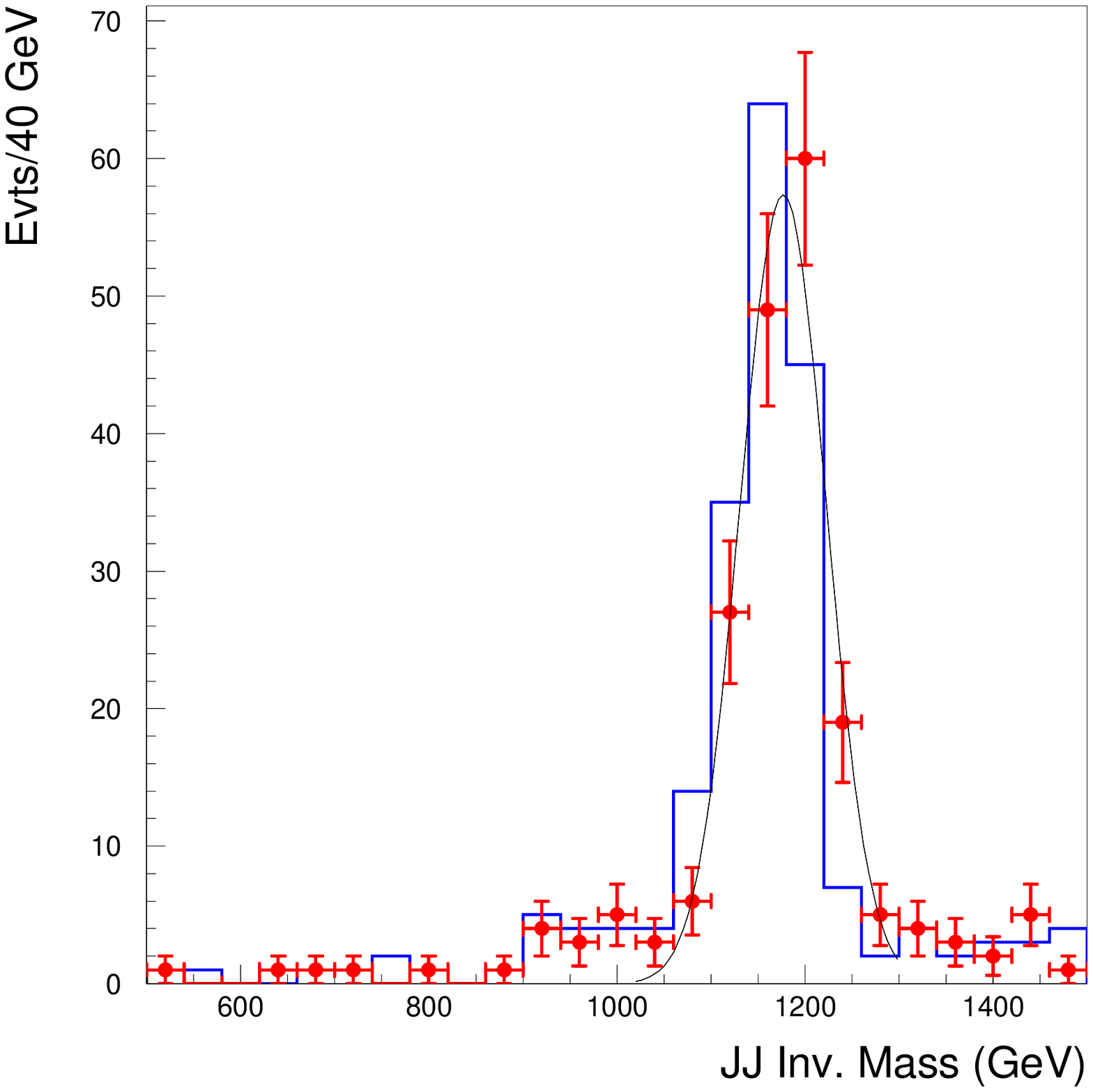,bbllx=0,bblly=0,bburx=565,bbury=565,width=7.5cm}
\caption{(Left) Decay angle distribution of the muons from 
 $G_3 (3200 \rm{GeV})\rightarrow \mu\mu$; (Right) Invariant jet-jet mass 
 of $G_1 (1200 \rm{GeV})$ 
produced in $G_3\rightarrow G_1G_1$ and
$G_1  \rightarrow$ jet jet. }
\label{fig:kk3}
\end{center}
\end{figure}

\begin{figure}[htbp]   
\begin{center} 
\epsfig{file=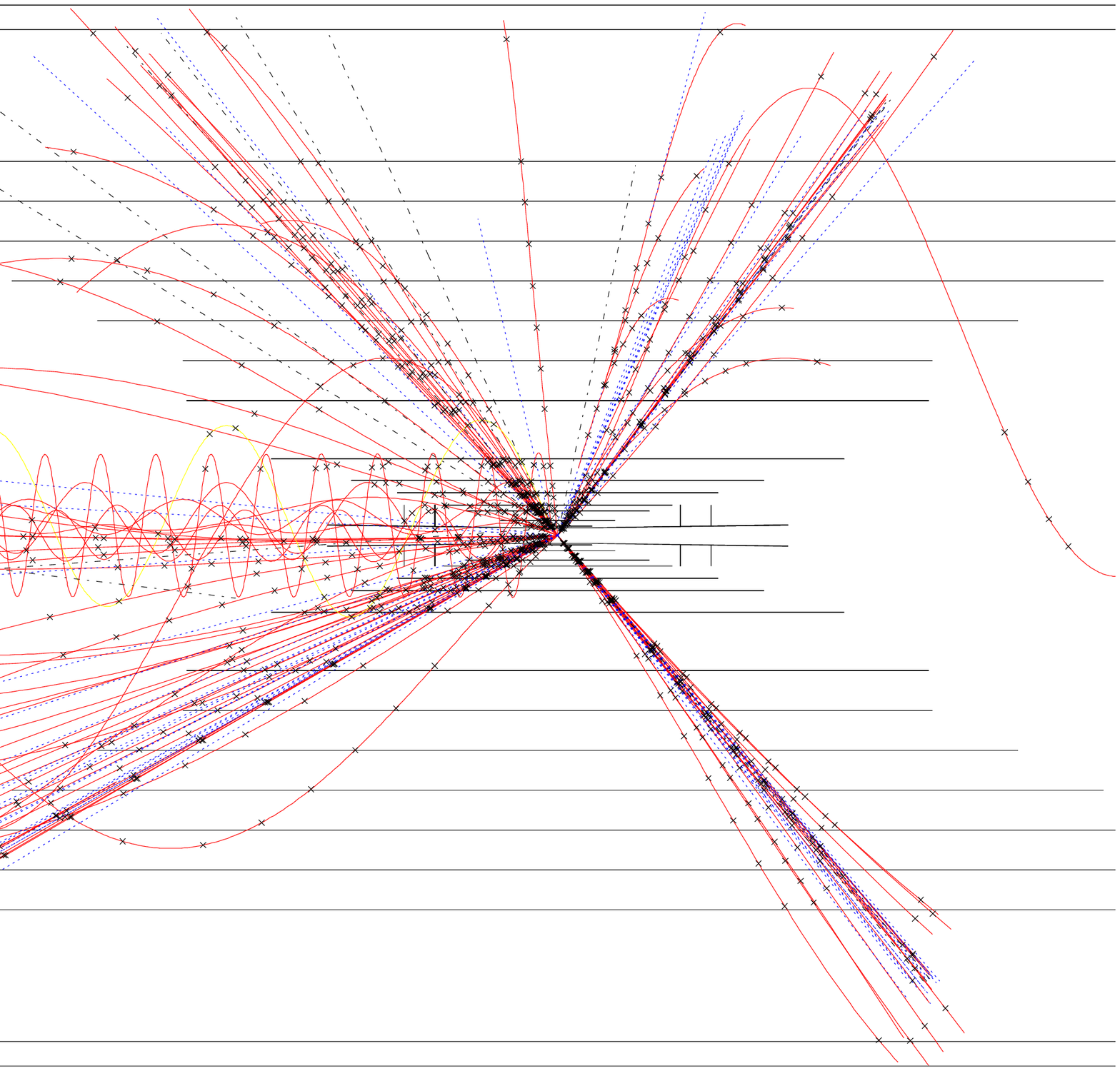,bbllx=0,bblly=20,bburx=600,bbury=700,width=6cm}
\epsfig{file=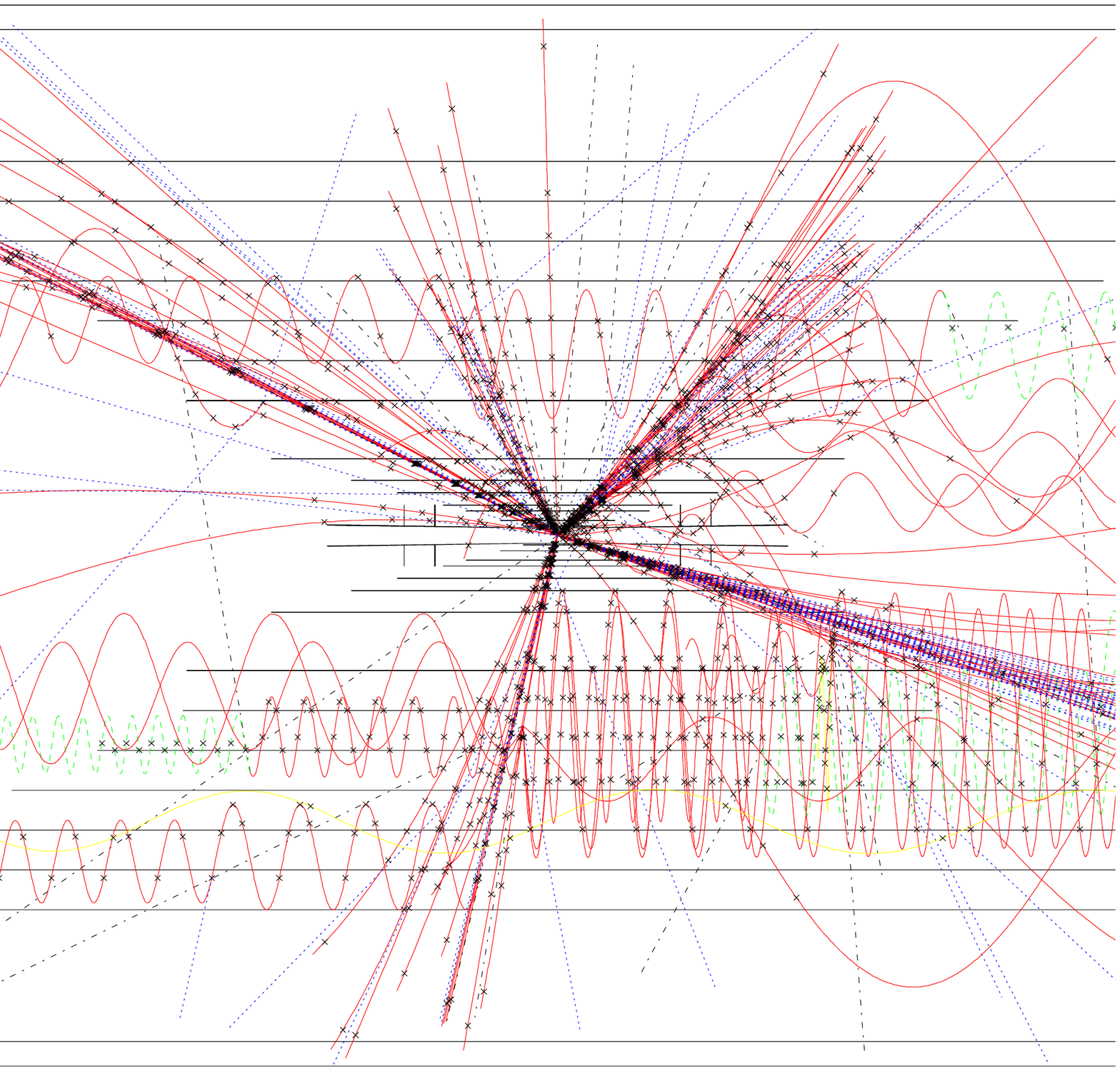,bbllx=0,bblly=20,bburx=600,bbury=700,width=6cm}
\caption{Two events in a CLIC central detector with decay
$G_3\rightarrow G_1G_1 \rightarrow$ four jets }
\label{fig:kk4}
\end{center}
\end{figure}

The resonance spectrum was chosen such that the first resonance $G_1$
has a mass around 1200 GeV, just outside the reach of a
TeV class LC, and consequently the mass of the third resonance 
$G_3$ will be around
3200 GeV, as shown in Fig.~\ref{fig:kk1}.
The CMS energy for the $e^+e^-$ collisions of CLIC was taken to be 3.2 TeV
in this study.
Mainly the muon and photon decay modes of the graviton have been studied.
The  events used  to reconstruct $G_3$ resonance signal were selected
either via 
two muons or two $\gamma$'s with $E> 1200$ GeV and $|\cos\theta| <0.97$.
Two typical events, one  decaying into muons and the other 
in photons, are shown in Fig.~\ref{fig:kk2}. 
The background from overlaid two photon events -- on average four events
per bunch crossing --is typically important only for angles below 120 mrad, 
i.e. outside the considered  signal search region.

First we study the precision with which we can measure the shape, i.e.
the  $c$ and  $M$ 
parameters, of  the observed  new resonance.
A scan similar to the one at the  $Z$ at LEP
was made for an integrated luminosity of  1 ab$^{-1}$.
An example of the cross section measurements and the $\chi^2$ fit
through the points for spectra generated with  different $c$ and  $M$ values
is shown in Fig.~\ref{fig:kk1}b.
The precision with which the cross sections are measured allows one to 
determine $c$ to  0.2\%, and $M$ to better than 0.1\%.

Next we determine some key  properties of the new resonance:
  the spin and the ratio of decay modes.
The graviton is a spin-two object. Fig.~\ref{fig:kk3} shows the 
decay angle of the  fermions $G\rightarrow \mu\mu$ 
for the $G_3$ graviton, using PYTHIA/SIMDET, for 1  ab$^{-1}$ of data,
 including CLIC machine background.
The typical spin-two structure of the decay angle 
 of the resonance is clearly visible.

For gravitons as proposed in~\cite{dhr,rizzo3} one expects
{$BR(G \rightarrow \gamma\gamma)/BR(G \rightarrow \mu\mu)$ = 2.}
With the present SIMDET simulation we get
efficiencies in mass peak ($\pm 200$  GeV) 
of 84\% and 97\% for detecting the muon and photon decay modes, respectively.
With cross 
sections of $O(pb)$, $\sigma_{\gamma\gamma}$ and $\sigma_{\mu\mu}$
can be determined to better than a per cent.
Hence the ratio $BR(G\rightarrow \gamma\gamma)/BR(G\rightarrow
\mu\mu)$ can be determined to an accuracy of 1\%  or better.

\begin{figure}[htp]   
\begin{center} 
\epsfig{file=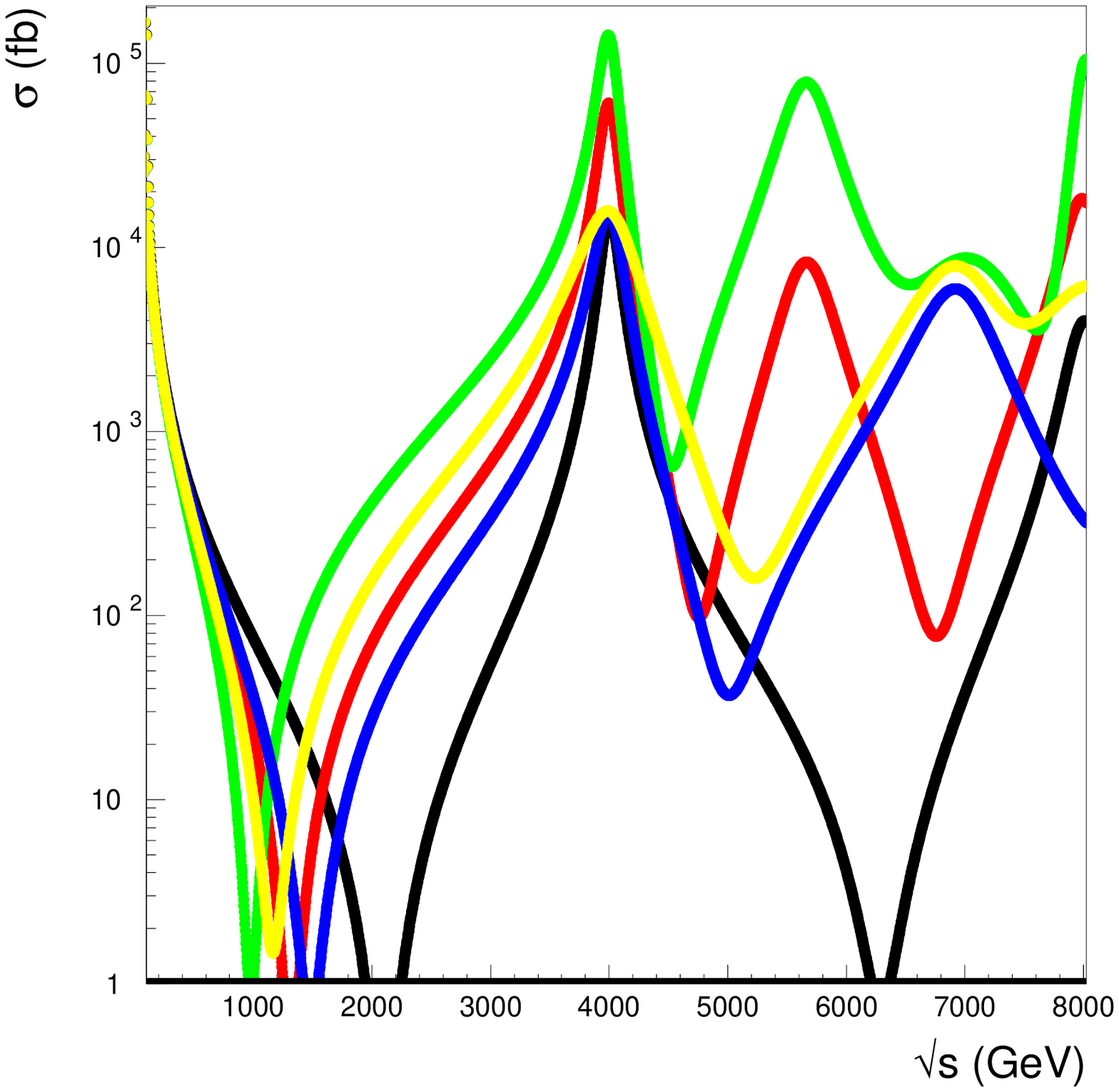,bbllx=30,bblly=20,bburx=540,bbury=520,width=5cm}
\epsfig{file=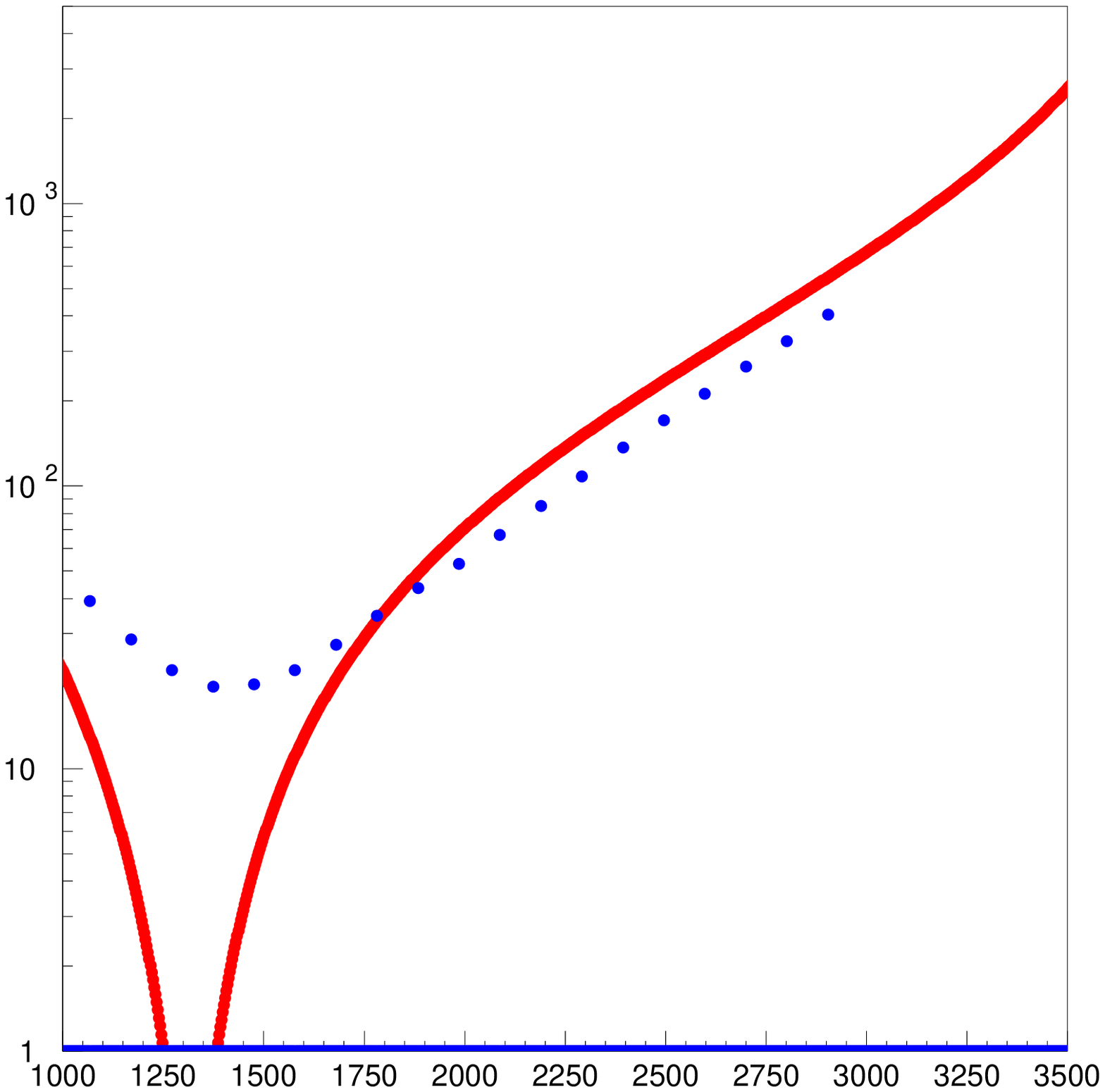,bbllx=30,bblly=20,bburx=540,bbury=520,width=5cm}
\epsfig{file=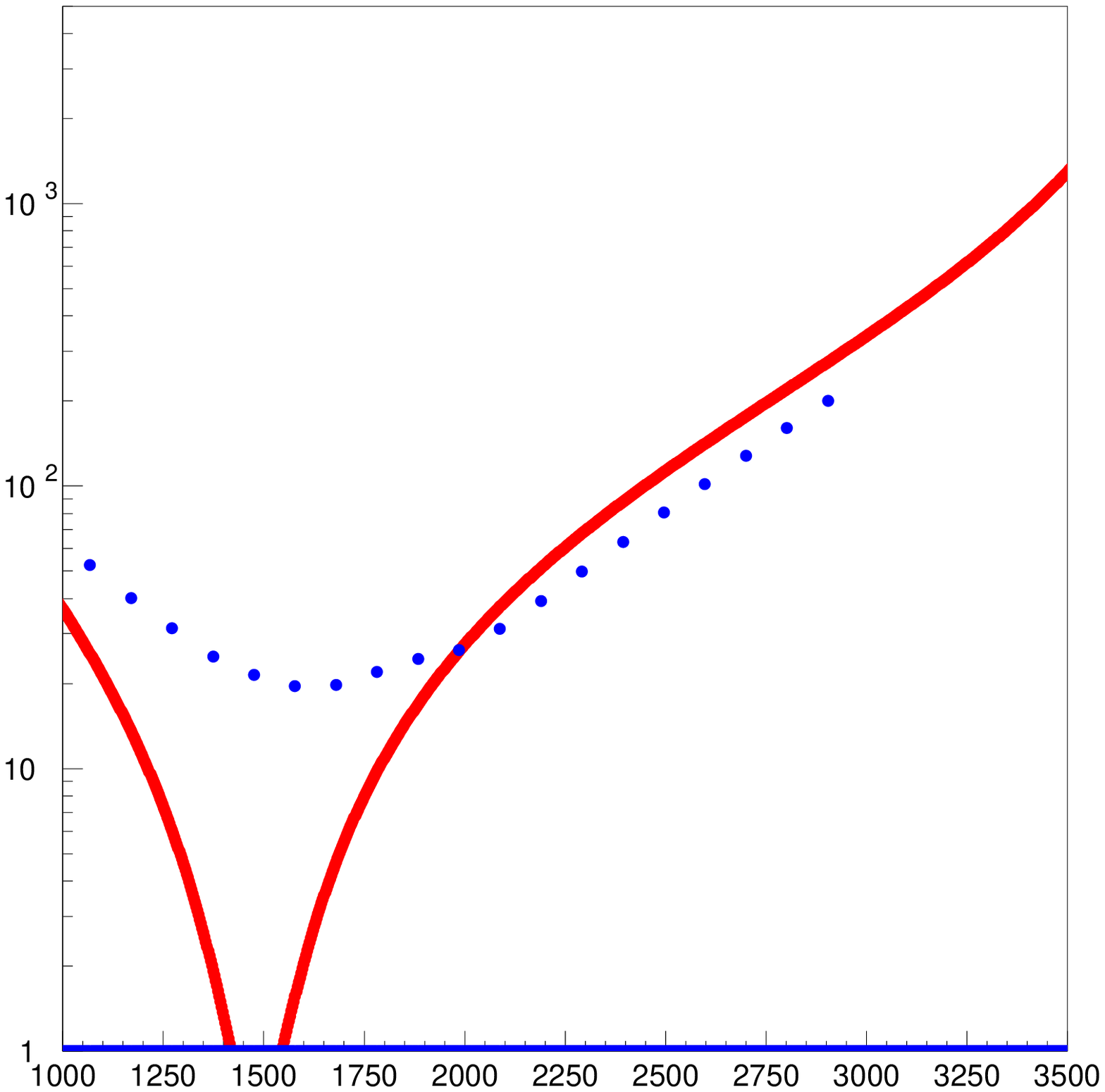,bbllx=30,bblly=20,bburx=540,bbury=520,width=5cm}
\caption{(Left) The $\sigma_{\mu\mu}$ cross section for
different models for TeV scale extra dimensions.
(Middle and Right) The $\sigma_{\mu\mu}$ cross section before (solid line)
and after (dashed line)  smearing by the CLIC luminosity spectrum, for 
two models.}
\label{fig:kk5}
\end{center}
\end{figure}

Finally, if the CMS energy of the collider is large 
enough to produce the first three resonances states, one has 
the intriguing possibility to measure the graviton 
self-coupling via the $G_3\rightarrow G_1G_1 $ decay~\cite{rizzo3}.
The dominant decay mode will be
$G_1 \rightarrow$ gluon-gluon or $q\bar q \rightarrow$ two jets.
Fig.~\ref{fig:kk4} shows the resulting spectacular event signature of
four jets of about 500 GeV each in the detector (no background 
is overlaid). These jets  can be used to reconstruct $G_1$.
Fig.~\ref{fig:kk3}b shows the  reconstructed $G_1$ invariant mass.
The histogram does not include  background while the data points include 
10 bunch crossings of background overlaid on the signal events.
Hence the mass of  $G_1$ can be well reconstructed and is not significantly
distorted by the $\gamma\gamma$ background. 

In summary a multi-TeV collider, such as 
 CLIC, will allow for a precise determination of the shape and of
mass of the new resonance(s), and of its spin.

\section{TeV scale models}

Another class of models, which leads to a resonance structure in the 
energy dependence of the two-fermion cross section, are those with a TeV scale 
extra dimension~\cite{anton}. 
In the simplest versions of these theories, only the 
SM gauge fields are in the bulk whereas the fermions remain at one of 
the two orbifold fixed points; Higgs fields may lie at the fixed points or 
propagate in the bulk. In such a model, to a good 
approximation, the masses of the KK tower states are given by $M_n=nM_c$, 
where $M_c=R_c^{-1}$ is the compactification scale, $R_c$ is the 
compactification radius. The mass of the lowest lying KK state is constrained 
to be rather large, i.e. a few TeV, due to bounds arising from precision 
measurements{\cite {big}}. 

The masses and couplings of the KK excitations are compactification scheme 
dependent and lead to a rather complex KK spectrum. Examples of models 
are shown in Fig.~\ref{fig:kk5}.
The  position of the peaks and the dips and their 
corresponding cross sections 
 and widths can be used to uniquely identify the 
extra-dimensional model. As an example  two of these models were taken 
and the production cross section was folded with the CLIC luminosity spectrum.
The results are shown in Fig.~\ref{fig:kk5} for two dip positions,
since the peaks are likely to be beyond the reach of a 3 TeV collider.
The structure of the dips is largely kept, but it is smeared out and 
somewhat systematically shifted. In any case also here 
CLIC data will be sensitive to the model parameters, and will allow 
to disentangle different scenarios

\section{Conclusions}

The direct production of KK excitations in the TeV range was studied for CLIC,
using examples of models based on RS and TeV scale extra dimensions. 
The backgrounds at CLIC and its smeared luminosity spectrum are 
not preventive to make precision measurements of the model parameters.
In particular, for the RS model it was shown that the key discriminating 
properties of these resonances can be reconstructed and the underlying
model parameters can be determined precisely. 
Hence if KK excitations appear in the two-fermion processes
cross sections in the TeV
range, CLIC will be an ideal tool to study in detail the properties of
these resonances.




\end{document}